\documentclass[a4paper,conference]{IEEEtran}
\usepackage{graphicx}

\usepackage{cite}

\ifCLASSINFOpdf
\else
\fi
\usepackage{amsmath}
\usepackage{algorithmic}
\usepackage{url}

\begin{document}
%
\title{Ensemble Reversible Data Hiding}

\author{Hanzhou Wu$^1$, Wei Wang$^1$, Jing Dong$^1$ and Hongxia Wang$^2$\\
$^1$Institute of Automation, Chinese Academy of Sciences, Beijing 100190, China\\
$^2$School of Information Science and Technology, Southwest Jiaotong University, Chengdu 611756, China\\
Email: h.wu.phd@ieee.org, \{wwang, jdong\}@nlpr.ia.ac.cn, hxwang@swjtu.edu.cn
}


%


\maketitle

\begin{abstract}
The conventional reversible data hiding (RDH) algorithms often consider the host as a whole to embed a secret payload. In order to achieve satisfactory rate-distortion performance, the secret bits are embedded into the noise-like component of the host such as prediction errors. From the rate-distortion optimization view, it may be not optimal since the data embedding units use the identical parameters. This motivates us to present a segmented data embedding strategy for efficient RDH in this paper, in which the raw host could be partitioned into multiple subhosts such that each one can freely optimize and use the data embedding parameters. Moreover, it enables us to apply different RDH algorithms within different subhosts, which is defined as ensemble. Notice that, the ensemble defined here is different from that in machine learning. Accordingly, the conventional operation corresponds to a special case of the proposed work. Since it is a general strategy, we combine some state-of-the-art algorithms to construct a new system using the proposed embedding strategy to evaluate the rate-distortion performance. Experimental results have shown that, the ensemble RDH system could outperform the original versions in most cases, which has shown the superiority and applicability.
\end{abstract}


%
\IEEEpeerreviewmaketitle

\section{Introduction}
Reversible data hiding (RDH) \cite{tian:de}, also called reversible watermarking, enables us to hide a payload into a host by slightly altering the host without introducing noticeable artifacts. And, for a receiver, both the original host content and the embedded information can be fully reconstructed. It is quite desirable in applications that require no degradation of the original content such as remote sensing and military. RDH is fragile, meaning that, when the marked host was manipulated, one will find it is not authentic and the original host may not be fully retrieved.

A number of RDH algorithms [2]-[12] have been reported, most of which use the prediction errors (PEs) of the cover elements to achieve RDH by histogram shifting (HS) \cite{ni:hs} or its variants \cite{sachnev:sp}. The prediction procedure enables us to produce a prediction error histogram (PEH). The HS operation allows us to reversibly embed a secret payload into the PEH. The conventional RDH algorithms often consider the host as a whole such that all PEH bins are processed with the identical parameters, which may be not optimal from the rate-distortion optimization view. This has been utilized in Li \emph{et al}'s work \cite{li:mhs}, which uses multiple histograms modification (MHM). As mentioned in their work, the data embedding capacity in a single layer is limited. Thus, the distortion may be high when the payload increases for a high layer embedding.

There are two reasonable explanations. First, their work uses a chessboard prediction pattern \cite{sachnev:sp}. It allows us to embed a half payload into the dot set. After that, the cross set can be used for embedding the other half payload. However, after embedding in the dot set, the cross set will use the altered pixels for prediction, which may reduce the prediction accuracy, resulting in degradation of the rate-distortion performance. Second, the two embedding channels use the same prediction and data embedding procedure. Actually, different predictors or data embedding algorithms have different performance on different subhosts. Data embedding in smooth region corresponds to better performance. It is quite desirable to apply such pixel prediction or embedding algorithms that they provide superior performance to a smooth subhost. For a complex subhost, other potential algorithms may be preferred. For example, in an image, different image blocks may have different local characteristics, which allows us to separately embed data into them by different RDH algorithms\footnote{Two RDH algorithms could be considered as different as long as the core steps are different, e.g., the predictor is different while the others are the same.}.

This has motivated us to present an ensemble data embedding strategy for RDH in this paper, in which the raw host is divided into multiple subhosts so that each subhost enables us to apply different RDH algorithms and separately use the optimized parameters. Experiments on public image dataset have demonstrated the superiority and applicability.

The rest of this paper are organized as follows. In Section II, we present the ensemble data embedding strategy, followed by a detailed scheme in Section III. Experiments on public dataset are provided and analyzed in Section IV. Finally, we conclude this paper in Section V.

\section{Ensemble Reversible Embedding}
Let $\textbf{X} = \{x_{i,j}\}^{h\times w}$ and $\textbf{O} = \{o_{i,j}\}^{h\times w}$ respectively denote the cover image to be embedded and its original version, where $x_{i,j},~o_{i,j} \in \{0, 1, ..., 2^d-1\}$, e.g., $d = 8$. $\textbf{X}\equiv\textbf{O}$ if $\textbf{X}$ was never embedded. We expect to embed a message $\textbf{m}\in \{0, 1\}^l$ into $\textbf{X}$ to generate such a marked image $\textbf{Y} = \{y_{i,j}\}^{h\times w}$ that the distortion between $\textbf{Y}$ and $\textbf{O}$, denoted by $D(\textbf{Y}, \textbf{O})$, is as low as possible. For compactness, we will sometimes say ``pixel $x_{i,j}$'' representing a pixel $x_{i,j}$ located at position $(i,j)$, whose value is $x_{i,j}$. And, $\textbf{X}$ represents a pixel-set containing all the pixels belonging to $\textbf{X}$.

We collect $N$ subhosts of $\textbf{X}$, i.e., $\textbf{X}_1$, $\textbf{X}_2$, ..., $\textbf{X}_N$, where $\cup_{i=1}^{N}\textbf{X}_i\subset \textbf{X}$. And, for some $i<j$, there may exist $\textbf{X}_i\cap\textbf{X}_j\neq\emptyset$. Notcie that, one can also set that the
subhosts are disjoint in practice. In RDH, we would like to use the noise-like component to carry a payload. For example, in prediction based RDH methods, a predictor is required to predict the cover elements to be embedded, by which the PEs can be obtained. We use $\textbf{e}_i = \{e_j^{(i)}\}_{j=1}^{N_i}, N_i\leq |\textbf{X}_i|$ to denote the noise-like vector of $\textbf{X}_i$. For compactness, we consider $N_i = |\textbf{X}_i|$ since one can always keep the non-embedded elements unchanged. We are to modify $\textbf{e}_i, 1\leq i\leq N$, to hide secret data. To better generalize it, we consider the embedding unit as a vector, rather than a single element (though this assumption will not be mentioned again). Namely, we divide each $\textbf{e}_i$ into $M_i$ disjoint vector-units, each of which is sized as $B_i$. Thus, we have $M_i\times B_i = N_i$, and rewrite $\textbf{e}_i = \{\textbf{b}_j\}_{j=1}^{M_i}$, where $\textbf{b}_j = \{e_k^{(i)}\}_{k=j\cdot B_i - B_i+1}^{j\cdot B_i}$. Notice that, $\textbf{e}_i$ may be further partitioned into sub-components for embedding.

We expect to find suitable RDH algorithms and embedding parameters to hide $\textbf{m} = \{\textbf{m}_1, \textbf{m}_2, ..., \textbf{m}_{N}\}$ to $\{\textbf{e}_1, \textbf{e}_2, ..., \textbf{e}_N\}$ separately so that the resulting overall distortion $D(\textbf{Y}, \textbf{O})$ can be as low as possible.

Let $\mathcal{A} = \{A_1, A_2, ..., A_{|\mathcal{A}|}\}$ and $\mathcal{P} = \{P_1, P_2, ..., P_{|\mathcal{P}|}\}$ be a set of candidate data embedding algorithms and a set including all possible parameter-sets. We expect to find such $\mathcal{R}_{\text{opt}}\in\mathcal{A}^{N}$ and $\mathcal{Q}_{\text{opt}}\in\mathcal{P}^{N}$ that all $\textbf{m}_i$ can be carried by $\textbf{e}_i$ and the overall distortion is as low as possible. Notice that, $\{\textbf{m}_1, \textbf{m}_2, ..., \textbf{m}_{N}\}$ will be orderly embedded into $\{\textbf{e}_1, \textbf{e}_2, ..., \textbf{e}_{N}\}$ with $\mathcal{R}_{\text{opt}}$ and $\mathcal{Q}_{\text{opt}}$. Namely, $\textbf{m}_i$ will be embedded into $\textbf{e}_i$ (corresponding to $\textbf{X}_i$) using $\mathcal{R}_{i}\in \mathcal{R}_{\text{opt}}$ and $\mathcal{Q}_{i}\in\mathcal{Q}_{\text{opt}}$ for $1\leq i\leq N$. For some $i<j$, if $\textbf{X}_i\cap\textbf{X}_j\neq\emptyset$, an element belonging to both sets may be modified twice. Thus, the data extraction procedure may be performed in an inverse manner.

Let
\begin{equation*}
\textbf{Y}^{(k)} = \text{Emb}(\textbf{X};\textbf{m}_{1\sim  k}, \textbf{e}_{1\sim k}, A_{i_1\sim i_k}, P_{j_1\sim j_k})
\end{equation*}
be the marked host by orderly embedding $\{\textbf{m}_1,$ $\textbf{m}_2,$ ..., $\textbf{m}_k\}$ into $\{\textbf{e}_1,$ $\textbf{e}_2,$ ..., $\textbf{e}_k\}$ using $\{A_{i_1},$ ..., $A_{i_k}\}$ and $\{P_{j_1},$ ..., $P_{j_k}\}$. Here, for all $1\leq t\leq k$, we have $i_t\in [1, |\mathcal{A}|]$ and $j_t\in [1, |\mathcal{P}|]$. There may exist some $a<b$ such that $i_a = i_b$ or $j_a = j_b$.

We use $[\textbf{Y}^{(k)}]$ to denote all the elements that were used to carry $\{\textbf{m}_1,$ $\textbf{m}_2,$ ..., $\textbf{m}_k\}$, namely, the elements in $[\textbf{Y}^{(k)}]$ are belonging to at least one of $\{\textbf{e}_1,$ $\textbf{e}_2,$ ..., $\textbf{e}_k\}$. Let $D(\textbf{Y}^{(k)},\textbf{O})$ denote the overall distortion between $\textbf{Y}^{(k)}$ and $\textbf{O}$. Therefore, our optimization task can be written as:
\begin{equation}
(\mathcal{R}_{\text{opt}}, \mathcal{Q}_{\text{opt}}) =
\underset{(\mathcal{R}, \mathcal{P})\in (\mathcal{A}^{N}, \mathcal{P}^{N})}
{\text{arg min}} D(\textbf{Y}^{(N)},\textbf{O}).
\end{equation}

We here limit ourselves to an additive distortion defined as:
\begin{equation}
D(\textbf{Y}, \textbf{O}) = \sum_{y_{i,j}\in \textbf{Y}}\rho_{i,j}(y_{i,j}, \textbf{O}),
\end{equation}
where $\rho_{i,j}:\{0, ..., 2^d - 1\}\times \{0, ..., 2^d - 1\}^{h\times w}\mapsto R$ exposes the cost of changing $o_{i,j}$ to $y_{i,j}$.
The additive assumption is reasonable since the interaction between two pixels that are far away from each other can be roughly ignored. And, the interaction between adjacent pixels can be captured by well designing $\rho$. A commonly used additive distortion measure is mean squared error (MSE), i.e.,
\begin{equation*}
D(\textbf{Y}, \textbf{O}) = \sum_{y_{i,j}\in \textbf{Y}}(y_{i,j} - o_{i,j})^2/\sum_{y_{i,j}\in \textbf{Y}} 1.
\end{equation*}

For compactness, we here use $D([\textbf{Y}^{(k)}],\textbf{O})$ to represent the distortion between $[\textbf{Y}^{(k)}]$ and $\textbf{O}$. Accordingly,
\begin{equation}
D(\textbf{Y}^{(k)}, \textbf{O}) = D([\textbf{Y}^{(k)}], \textbf{O}) + D(\textbf{Y}^{(k)}-[\textbf{Y}^{(k)}], \textbf{O}).
\end{equation}

Since we embed $\textbf{m}_k$ into $\textbf{e}_k$ only when $\{\textbf{m}_1, \textbf{m}_2, ..., \textbf{m}_{k-1}\}$ have been previously embedded into $\{\textbf{e}_1, \textbf{e}_2, ..., \textbf{e}_{k-1}\}$ respectively, we have
\begin{equation}
\begin{split}
&D([\textbf{Y}^{(k)}], \textbf{O}) = \sum_{y_{i,j}^{(k)}\in [\textbf{Y}^{(k)}]}\rho_{i,j}(y_{i,j}^{(k)}, \textbf{O})\\
& = \sum_{y_{i,j}^{(k)}\in [\textbf{Y}^{(k-1)}]}\rho_{i,j}(y_{i,j}^{(k)}, \textbf{O}) + \sum_{y_{i,j}^{(k)}\notin[\textbf{Y}^{(k-1)}]}\rho_{i,j}(y_{i,j}^{(k)}, \mathcal{O})\\
& = D([\textbf{Y}^{(k-1)}], \textbf{O}) + D([\textbf{Y}^{(k)}] - [\textbf{Y}^{(k-1)}],\textbf{O}),
\end{split}
\end{equation}
where $[\textbf{Y}^{(k)}] - [\textbf{Y}^{(k-1)}]$ corresponds to a \emph{subset} of $\textbf{X}_k$ since there may be $\textbf{X}_k\cap\textbf{X}_j\neq\emptyset$ for some $j<k$. We rewrite:
\begin{equation}
\begin{split}
D(\textbf{Y}^{(k)}, \textbf{O}) &= D([\textbf{Y}^{(k-1)}], \textbf{O}) + D(\textbf{Y}^{(k)}-[\textbf{Y}^{(k)}], \textbf{O})\\
&~~~+ D([\textbf{Y}^{(k)}] - [\textbf{Y}^{(k-1)}],\textbf{O}).
\end{split}
\end{equation}

$D(\textbf{Y}^{(k)}-[\textbf{Y}^{(k)}], \textbf{O})$ can be determined before we compute $D(\textbf{Y}^{(k)}, \textbf{O})$. $D([\textbf{Y}^{(k-1)}], \textbf{O})$ and $D([\textbf{Y}^{(k)}] - [\textbf{Y}^{(k-1)}],\textbf{O})$ rely on the data embedding algorithms and parameters. $(\mathcal{A}_{\text{opt}}, \mathcal{P}_{\text{opt}})$ corresponds to a minimized $D(\textbf{Y}^{(N)},\textbf{O})$, i.e., $D_\text{min}(\textbf{Y}^{(N)},\textbf{O})$.

We write:
\begin{equation}
\begin{split}
&D_\text{min}(\textbf{Y}^{(k)}, \textbf{O}) = \underset{A_{i_1\sim i_k},P_{j_1\sim j_k}}{\text{min}}~~\bigg(D([\textbf{Y}^{(k-1)}], \textbf{O}) + \\
&~~~~D([\textbf{Y}^{(k)}] - [\textbf{Y}^{(k-1)}],\textbf{O}) \bigg) + D(\textbf{Y}^{(k)}-[\textbf{Y}^{(k)}], \textbf{O})\\
&\approx D(\textbf{Y}^{(k)}-[\textbf{Y}^{(k)}], \textbf{O}) + D_\text{min}([\textbf{Y}^{(k-1)}], \textbf{O}) + \\
&~~~~~~~~~~~~~~~\underset{A_{i_k},P_{j_k}}{\text{min}}~\epsilon_k\cdot D([\textbf{Y}^{(k)}] - [\textbf{Y}^{(k-1)}],\textbf{O}),
\end{split}
\end{equation}
where
\begin{equation}
\begin{split}
&D_\text{min}([\textbf{Y}^{(k-1)}], \textbf{O}) \approx D_\text{min}([\textbf{Y}^{(k-2)}], \textbf{O})\\
&~~~~~~+\underset{A_{i_{k-1}},P_{j_{k-1}}}{\text{min}}~\epsilon_{k-1}\cdot D([\textbf{Y}^{(k-1)}] - [\textbf{Y}^{(k-2)}],\textbf{O}),
\end{split}
\end{equation}
and
\begin{equation}
\begin{split}
D_\text{min}([\textbf{Y}^{(1)}], \textbf{O})=
\underset{A_{i_1},P_{j_1}}{\text{min}}~D([\textbf{Y}^{(1)}],\textbf{O}),
\end{split}
\end{equation}
where $\{\epsilon_2, ..., \epsilon_k\}$ are empirical factors. Actually, for $i>1$, $\epsilon_i$ reflects the interaction between $[\textbf{Y}^{(i-1)}]$ and $\textbf{X}_i$. From an implementation view, the determination of $\{\epsilon_2, ..., \epsilon_k\}$ can be avoided. That is, in practice, one can determine the overall distortion excluding $\textbf{X}_i$ previously and then focus on optimizing the embedding algorithm and parameters for $\textbf{X}_i$ since the subhosts are orderly embedded\footnote{It is also due to the ``additive'' distortion and inclusion-exclusion principle.}. Thereafter, the distortion introduced by $\textbf{X}_i$ can be easily determined.

Eqs. (6-8) show the state-transition equations to determine $D_\text{min}(\textbf{Y}^{(N)},\textbf{O})$. It essentially breaks the distortion optimization down into a sequence of subproblems. During the distortion optimization process, one can construct $\mathcal{R}_{\text{opt}}$ and $\mathcal{Q}_{\text{opt}}$, which are used for data embedding simultaneously. It is seen that, most conventional algorithms use $N = 1$, which is a special case of our model. The time complexity to enumerate all $(\mathcal{R}, \mathcal{P})$ for Eq. (1) is $O(|\mathcal{A}|^N\cdot |\mathcal{P}|^N)$. By applying Eqs. (6-8), it is significantly reduced as $O(N\cdot |\mathcal{A}| \cdot |{\mathcal{P}}|)$. By using a small number of candidate RDH algorithms and heuristically optimizing the data embedding parameters, the enumeration complexity could further decline significantly.
\begin{figure*}[!t]
\centering
\includegraphics[width=7in]{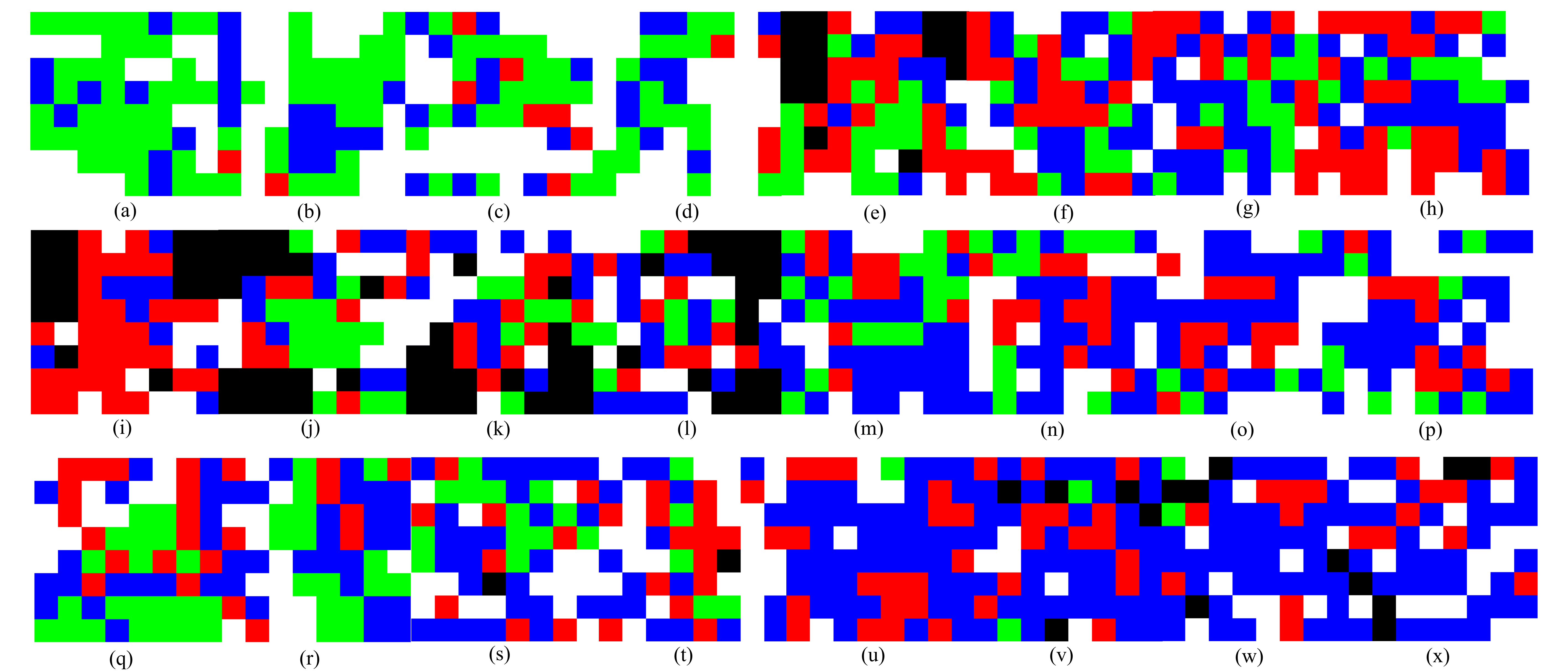}
\caption{The RDH algorithm distribution maps of the proposed ensemble algorithm for the $5^{\text{th}}\sim 8^{\text{th}}$-layer embedding (i.e., the multiple embedding strategy was applied) for each test image: (a$\sim$d) Airplane, (e$\sim$h) Lena, (i$\sim$l) Baboon, (m$\sim$p) Lake, (q$\sim$t) Boat and (u$\sim$x) Peppers. Here, we use $N = 64$, and the border area, i.e., $\textbf{X}_\text{aux}$, is ignored for better presentation. The red, green, blue, white areas are representing Wu \emph{et al}'s method, Li \emph{et al}'s method, Hsu \emph{et al}'s method and Ni \emph{et al}'s method, respectively. Moreover, the black blocks indicate that they cannot carry additional data, i.e., no RDH algorithm was applied. The input cover image of a higher layer embedding is a roatated version (a degree of 90) of the marked image obtained from the previous-layer embedding. It means that, the maps shown here have been rotated, e.g., (a) is rotated by a degree of 360 (corresponding to the original image, not that in the previous layer), (b) is rotated by a degree of 450, (c) and (d) for a degree of 540 and 630, respectively. Additionally, since we fix $N=64$ here, the corresponding rate-distortion performance may be not optimal for the above maps.}
\end{figure*}

\section{Detailed Ensemble RDH Scheme}
We present a detailed scheme in this section. However, it is always free for us to design any new ensemble scheme. We divide $\textbf{X}$ to two subsets, denoted by $\textbf{X}_\text{emb}$ and $\textbf{X}_\text{aux}$, i.e.,
\begin{equation*}
\textbf{X}_\text{emb} = \{x_{i,j} | i\in (T, h - T], j\in (T, w-T]\},
\end{equation*}
and
\begin{equation*}
\textbf{X}_\text{aux} = \textbf{X}\setminus \textbf{X}_\text{emb},
\end{equation*}
where $T$ is a system parameter and a pixel position is indexed from (1, 1) to $(h, w)$. $\textbf{X}_\text{aux}$ will be used to store auxiliary data such as the secret key. $\textbf{X}_\text{emb}$ will carry $\textbf{m}$ and other side information such as the location map. We segment $\textbf{X}_\text{emb}$ into $N$ image blocks from up to down and left to right, denoted by $\textbf{X}_\text{emb}^{(1)}$, $\textbf{X}_\text{emb}^{(2)}$, ..., $\textbf{X}_\text{emb}^{(N)}$. Here, $\cup_{i=1}^{N}\textbf{X}_\text{emb}^{(i)}\subset\textbf{X}_\text{emb}$, and for all $1\leq i < j\leq N$, $\textbf{X}_\text{emb}^{(i)}\cap\textbf{X}_\text{emb}^{(j)}=\emptyset$. One may consider that, for all $1\leq i\leq N$, we have $\textbf{X}_i = \textbf{X}_\text{aux}\cup\textbf{X}_\text{emb}^{(i)}$. Thus, we have $\textbf{X}_i\cap \textbf{X}_j = \textbf{X}_\text{aux}$ for all $1\leq i<j\leq N$. For simplicity, we assume that $|\textbf{X}_\text{emb}^{(i)}| \approx q^2$ for all $1\leq i\leq N$, where $q > 0$ is a positive integer. Therefore, we have $N \leq \left \lfloor \frac{h-2T}{q}\right \rfloor\cdot\left\lfloor\frac{w-2T}{q} \right \rfloor$.

We expect to find suitable RDH algorithms and parameters to embed $\textbf{m} = \{\textbf{m}_1, \textbf{m}_2, ..., \textbf{m}_{N}\}$ into $\{\textbf{X}_\text{emb}^{(1)}, \textbf{X}_\text{emb}^{(2)}, ..., \textbf{X}_\text{emb}^{(N)}\}$. And, some auxiliary data for data extraction and image recovery are embedded into $\textbf{X}_\text{aux}$. Since HS has been a most popular operation in RDH, we will use HS-based RDH algorithms for experiments. In detail, four state-of-the-art algorithms, i.e., Ni \emph{et al.} \cite{ni:hs}, Li \emph{et al.} \cite{li:mhs}, Wu \emph{et al.} \cite{hzwu:PPE} and Hsu \emph{et al.} \cite{hsu:smp}, are used to construct an ensemble RDH system. There are three reasons for why we select the four algorithms. First, they are HS based algorithms providing superior rate-distortion performance, and are relatively easy to simulate. Second, they allow us to use two pairs of peak-zero points at a time to embed the secret data. It is convenient for us to use them to build a new RDH system and optimize the embedding parameters. Third, a core contribution of these works is the pixel prediction procedure. Clearly, the prediction value of a cover pixel in \cite{ni:hs} can be fixed as zero all the time. In Li \emph{et al.}'s work, they use the mean value of neighbors to estimate the current pixel. In Wu \emph{et al.}'s work, a second-order prediction procedure is applied. And, in Hsu \emph{et al.}'s work, side corrections are considered. These different predictors enable us to better choose a suitable RDH algorithm for a subhost during the optimization process, which could benefit the overall rate-distortion performance. However, it is still pointed that, it is always free for a data hider to choose the candidate RDH algorithms. We cannot guarantee that our choice is optimal. One may take into account more candidate algorithms, which requires a higher computational cost.

\subsection{Preprocessing}
We use a key $K_p$ to produce a permutation of $\{1, 2, ..., N\}$, denoted by $\{p_1, p_2, ..., p_N\}$, to control data embedding order. It means that, we will first embed $\textbf{m}_{1}$ into $\textbf{X}_\text{emb}^{(p_1)}$. Then, we embed $\textbf{m}_{2}$ into $\textbf{X}_\text{emb}^{(p_2)}$ and so on. $K_p$ will be self-embedded into some pixels in $\textbf{X}_\text{aux}$ by LSB replacement, where the original LSBs of the specified pixels will be considered as a part of $\textbf{m}$. We have to self-embed $N$ and $T$ in advance as well. The process is similar to $K_p$. In default, we can set $T = 1$. The size of side information for storing $K_p$, $N$ and $T$ will have ignorable impact on the pure embedding payload. Since $\textbf{X}_\text{aux}$ is slightly modified, its impact on the overall distortion can be always roughly ignored during optimization.

\subsection{Data Embedding}
Assuming that, $\{\textbf{m}_{{1}}, \textbf{m}_{{2}}, ..., \textbf{m}_{{k-1}}\}$ have been previously carried by $\{\textbf{X}_\text{emb}^{(p_{1})}, \textbf{X}_\text{emb}^{(p_{2})}, ..., \textbf{X}_\text{emb}^{(p_{k-1})}\}$, we are to embed $\textbf{m}_{{k}}$. We apply the four RDH algorithms mentioned above to $\textbf{X}_\text{emb}^{(p_{k})}$. The RDH algorithm that results in the lowest overall distortion will be selected as the final algorithm for $\textbf{X}_\text{emb}^{(p_k)}$. It is noted that, for each RDH algorithm, during data embedding optimization, the side information involves three aspects, i.e., the location map $L_{p_k}$, the data-embedding parameters $E_{p_k}$ and the secret key $K_{p_k}$. $L_{p_k}$ is constructed to avoid underflow and overflow problem, which will be embedded into $\textbf{X}_\text{emb}^{(p_k)}$ together with $\textbf{m}_{{k}}$. Meanwhile, $E_{p_k}$ and $K_{p_k}$ should be embedded into $\textbf{X}_\text{aux}$ by LSB replacement. The LSBs of the specified pixels of $\textbf{X}_\text{aux}$ will be embedded into $\textbf{X}_\text{emb}^{(p_k)}$ as well.

In addition to $E_{p_k}$ and $K_{p_k}$, we need to embed the index of the selected RDH algorithm into $\textbf{X}_\text{aux}$ by a similar way. As mentioned above, since $L_{p_k}$ (which may be losslessly compressed in advance) should be embedded into $\textbf{X}_\text{emb}^{(p_k)}$, it may limit the pure embedding capacity of $\textbf{X}_\text{emb}^{(p_k)}$. It indicates that, it is possible that, the four algorithms are all non-embeddable, i.e., they all cannot carry $\textbf{m}_k$. In this case, we will not embed secret bits into $\textbf{X}_\text{emb}^{(p_k)}$ including the side information. It also means that, $|\textbf{m}_k| = 0$. Thus, we need another bit to tell a decoder whether the present subhost is embedded or not. The bit will be embedded into $\textbf{X}_\text{aux}$ as well. In case that the LSBs of $\textbf{X}_\text{aux}$ are not enough to carry the auxiliary data, one can use the second-LSB-plane, for which the original bits should be recorded and embedded into $\textbf{X}_\text{emb}$ as well. There has another suitable way to deal with the above problem. Namely, we replace a part of the LSBs of $\textbf{X}_\text{aux}$ with the auxiliary data for the present subhost. The modified LSBs of $\textbf{X}_\text{aux}$ will be recorded and embedded into the next subhost\footnote{The pure data embedding rate should exclude those bits for recovering the original LSBs of $\textbf{X}_\text{aux}$.}. After processing $\textbf{X}_\text{emb}^{(p_k)}$, we continue to process $\textbf{X}_\text{emb}^{(p_{k+1})}$ until $\textbf{m}$ and the necessary auxiliary data are all completely embedded.

\subsection{Data Extraction and Image Recovery}
For a data receiver, he first extracts $K_p$ and identifies data embedding order of the encoder side. Notice that, it is required that the LSBs for storing $K_p$ at the encoder side should not be overridden. Then, the receiver extracts $\textbf{m}$ in an inverse manner. Namely, he will first extract $\textbf{m}_N$ from the marked $\textbf{X}_\text{emb}^{(p_{N})}$, then extract $\textbf{m}_{N-1}$ from the marked $\textbf{X}_\text{emb}^{(p_{N-1})}$, and so on.
\begin{figure*}[!t]
\centering
\includegraphics[width=6.4in]{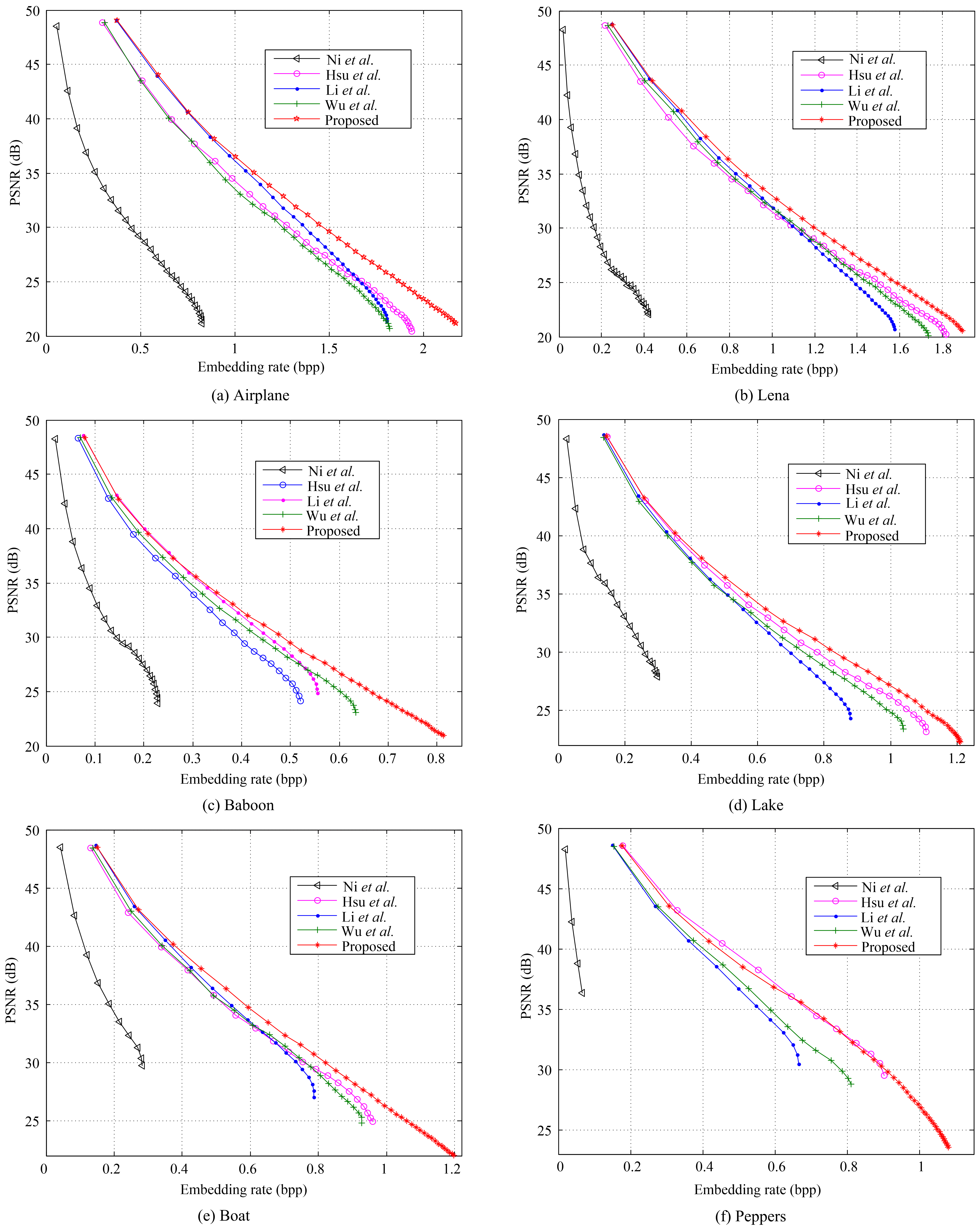}
\caption{The rate-distortion performance comparison between the state-of-the-art RDH algorithms introduced by Ni \emph{et al}. \cite{ni:hs}, Hsu \emph{et al}. \cite{hsu:smp}, Li \emph{et al}. \cite{li:mhs}, Wu \emph{et al}. \cite{hzwu:PPE}, and the proposed ensemble algorithm for the six standard test images. We here vary $m$ from 2 to 8 for rate-distortion optimization during each-layer embedding ($N = m^2$). The multiple embedding strategy was utilized to carry more data.
}
\end{figure*}

Accordingly, $\textbf{m}$ can be finally reconstructed. Meanwhile, the side information such as the location maps and the original LSBs of $\textbf{X}_\text{aux}$ can be correctly recovered as well. This allows the original image to be perfectly reconstructed.

It is mentioned that, one may use the multiple embedding strategy to embed as many message bits as possible, i.e., an original image may be embedded multiple times at the encoder side. In this case, the data extraction and image recovery process can be performed in a similar way. It is inferred that, the multiple embedding strategy ensures reversibility.

\section{Performance Evaluation and Analysis}
In this section, we will present experiments on public image dataset for performance evaluation and analysis. As mentioned above, four state-of-the-art RDH algorithms introduced by Ni \emph{et al.} \cite{ni:hs}, Li \emph{et al.} \cite{li:mhs}, Wu \emph{et al.} \cite{hzwu:PPE} and Hsu \emph{et al.} \cite{hsu:smp}, are used to construct a new ensemble RDH system. For each candidate algorithm, we introduce the important configuration in our simulation below.

For Ni \emph{et al.}'s algorithm, we generate the histogram directly from the cover image. Two pairs of peak-zero histogram bins are selected out for data embedding with the HS operation. For the two zero-bins, we record their positions and construct a location map losslessly compressed by arithmetic coding. It will be embedded into the cover image. The original location map corresponds to a binary map with the same size of the image, where ``1''s represent the pixels of zero-bins and ``0''s for the others. Notice that, it is possible that the number of ``1'' is zero. In addition, we assume that the absolute difference value of a peak-bin and the corresponding zero-bin should be no less than 2, which is to avoid extraction ambiguous.

For the algorithms introduced in Li \emph{et al.}, Wu \emph{et al.} and Hsu \emph{et al.}, the processes are similar. First of all, the boundary pixels are adjusted into the reliable range and recorded to constitute a location map compressed by arithmetic coding. Then, the pixels to be embedded are predicted according to the corresponding predictor. Thereafter, the secret data and auxiliary data are embedded by the corresponding operation. Since the candidate algorithms use prediction-errors and HS, we embed the secret data by using two pairs of peak-zero bins\footnote{One has to record the occurrences of zero-bins in the PEH for reversibility even though the number of occurrences of zero-bins is often zero.}, which is the same as the original ones.

In our simulation, we consider $N$ as a square number, i.e., $N$ = $m^2$, and set $T = 1$ in default. For a given image, we change $m$ from 2 to 8 by a step value of 1 for optimization. The best rate-distortion performance will be considered as the result since the data sender always has the freedom to choose better parameters. The multi-layer embedding strategy is applied to all candidate algorithms and the proposed ensemble system. Notice that, when to use the multi-layer embedding strategy for the proposed ensemble system, the optimized value of $N$ may be different for each embedding layer. As mentioned previously, there may exist some $\mathcal{X}_\text{emb}^{(i)}$ that it cannot carry additional data. To deal with this problem, after a single-layer embedding, the entire marked image will be rotated by a degree of 90 such that the new pixel-blocks may be different (if we divide $\textbf{X}_\text{emb}$ from left to right and from up to down). Notice that, in real world, there has no need to directly rotate the image, but to change the way for block division. Furthermore, in Li \emph{et al.}'s work, they divide a PEH into 16 sub-PEHs in default. In the new ensemble system, for all possible $\textbf{X}_\text{emb}^{(i)}$, when to use their work, we use only one sub-PEH, which can reduce the size of side information.

We take six standard test images\footnote{Downloaded at \url{http://sipi.usc.edu/database/database.php?volume=misc}} \emph{Airplane}, \emph{Lena}, \emph{Baboon}, \emph{Lake}, \emph{Boat} and \emph{Peppers} with a size of $512\times 512\times 8$ from smooth to complex for rate-distortion performance evaluation. The peak signal-to-noise ratio (PSNR, dB) is determined to evaluate the marked image quality. We focus on the pure data embedding rate, which does not include side information such as the location map. Fig. 1 shows the algorithm distribution maps for the $5^{\text{th}}\sim8^{\text{th}}$-layer embedding for each test image by the proposed ensemble system. It can be seen that, different image blocks (corresponding to different subhosts) have different characteristics and therefore use different candidate algorithms, which has demonstrated the applicability of the segmented and ensemble strategy.

Fig. 2 shows the rate-distortion performance comparison between the candidate algorithms and the new ensemble system. It is observed that, the new system can not only provide a relatively higher pure data embedding capacity, but also introduce a relatively lower distortion. It indicates that the proposed work has the potential to significantly improve the rate-distortion performance, which demonstrates the superiority. It is also observed that, for the \emph{Peppers} image, the PSNRs of our ensemble system are slightly lower than Hsu \emph{et al.}'s method for relatively low embedding rates. This indicates that, the candidate algorithms may not fully benefit the ensemble system, indicating that, we could consider more candidate RDH algorithms. It also implies that, the used image-block division method may not well exploit the statistical characteristics of the covers, which leads us to design more efficient ensemble system in the future.

\section{Conclusion and Discussion}
A number of RDH algorithms have been reported in the past twenty years. They have moved the field ahead rapidly. Most of them consider the cover as a whole for data embedding. Actually, different subhosts of the cover may have different characteristics, which implies that, we may be able to apply different data embedding algorithms to different subhosts so that better performance can be achieved. This paper presents a novel segmented and ensemble embedding strategy for RDH, which can deal with the above requirement. We also present a detailed ensemble RDH scheme. Experimental results have shown that, the proposed work could significantly improve the performance. The proposed work may have potential in RDH.

In addition, the ensemble perspective may be applicable to other subfields of information hiding such as steganography \cite{hzwu:gmba}. In this sense, one may redefine the ``ensemble'' term. And, a core research topic may be to formulate the optimization problem and find the optimal solution.

\section*{Acknowledgment}
This work was partly supported by NSFC (No. 61502496, U1536120, U1636201, U1736119, and 61772529) and the National Key Research and Development Program of China (No. 2016YFB1001003). It was also partly supported by the Key Lab of Information Network Security and the Ministry of Public Security of China.



%

\end{document}